# Special class of self-similar dynamics for Rayleigh-Taylor mixing with variable acceleration

Snezhana I. Abarzhi

The University of Western Australia, AUS

snezhana.abarzhi@gmail.com

Rayleigh-Taylor (RT) mixing has critical importance for a broad range of process in nature and technology, from supernovae and plasma fusion to oil recovery and nano-fabrication. In most instances, RT flows are driven by variable acceleration, whereas the bulk of existing studies have considered only constant and impulsive accelerations. By analyzing symmetries of RT dynamics for certain patterns of variable acceleration, we discover a special class of self-similar solutions and identify their scaling, correlations and spectra. We find that dynamics of RT mixing can vary from super-ballistic to sub-diffusive depending on the acceleration and retain memory of deterministic conditions for any acceleration. These rich dynamic properties considerably impact the understanding and control of RT relevant phenomena in fluids, plasmas, materials, and reveal conditions at which turbulence can be realized in RT mixing.

Significance statement:

Rayleigh-Taylor material mixing is common to occur in fluids, plasmas, materials, and has critical importance for a broad range of processes from celestial events to atoms. Examples include supernovae, planetary convection, Earth ionosphere, plasma fusion, reactive fluids, fossil fuel industry, optical communications, nano-fabrication. To advance knowledge of non-equilibrium dynamics and positively impact energy and environment, reliable theory of Rayleigh-Taylor mixing is in demand. We report the discovery of the special class of self-similar solutions for Rayleigh-Taylor mixing with variable acceleration. The dynamics are rich, varying from super-ballistics with high order to sub-diffusion with localized fluctuations, and retaining memory of deterministic conditions. We reveal dramatic impact these properties have on the understanding and control of Rayleigh-Taylor-relevant phenomena in nature and technology.

**Section - Introduction**

Turbulence is often considered as the only remaining problem of classical physics [1-3]. Its complexity and similarity fascinate scientists and mathematicians, demand due attention of engineers and practitioners, and inspire philosophers and poets [1-3]. Isotropy, homogeneity, locality - are the fundamental hypotheses advanced our understanding of turbulent dynamics [2,4]. Still the problem sustains the efforts applied, since realistic processes often depart from idealistic conditions [5]. Self-similar Rayleigh-Taylor (RT) mixing is a generic problem we inevitably encounter when expanding our knowledge of turbulence beyond canonical considerations [5].

Rayleigh-Taylor mixing is an intensive interfacial mixing developing when fluids of different densities are accelerated against their density gradients [6-8]. It occurs in nature and technology, at astrophysical and at molecular scales, under conditions of high and low energy densities [8]. Supernovae, fusion, oil recovery, nano-fabrication - are examples of processes, to which RT mixing is critically relevant [5]. In many instances, RT dynamics are driven by variable accelerations, whereas the bulk of existing studies have considered only constant and impulsive (e.g., shock-driven) accelerations. The former is referred as to classical RT instability (RTI); the latter - as to Richtmyer-Meshkov instability (RMI) [6-10]. This work analyzes symmetries and invariants of RT mixing driven by accelerations with power-law time dependence. We discover a special class of self-similar solutions and identify their scaling, correlations and spectra. We find that dynamics of RT mixing can vary from super-ballistic to sub-diffusion depending on the exponent of the acceleration's power-law, and retains memory of deterministic conditions for any exponent's value.

RT flows with variable acceleration are common to occur in fluids, plasmas, materials [5,8]. These include: RTI quenching ignition in inertial confinement fusion; blast wave driven RT mixing in supernova enabling conditions for synthesis of heavy mass elements; RT unstable plasma irregularities in the Earth ionosphere resulting in climate change on regional scales; RT mixing of water and oil limiting efficiency of oil recovery; RTI governing material transformation under impact in nano-fabrication [5,11-14]. In these vastly different physical regimes, RT flows exhibit some similar features of the evolution [8]. RTI starts to develop when the fluid interface is slightly perturbed near the equilibrium state [6]. The flow transitions from an initial stage, where the perturbation amplitude grows quickly, to a nonlinear stage, where the growth-rate slows and the interface is transformed to a composition of small-scale shear driven vortical structures and a large-scale coherent structure. Final stage is intensive interfacial mixing, whose dynamics is believed to be self-similar [6-10,15].

RT mixing is heterogeneous anisotropic process with non-local interactions among the many scales. It is an extreme challenge to study in its direct manifestations [5,8,15]. In experiments, the transient character and sensitivity of large scale dynamics to small scales and deterministic conditions impose tight requirements on the flow implementation, diagnostics and control [16-21]. In simulations, the needs to accurately track unstable interfaces and capture small-scale processes, the singular nature of numerical solutions and their dependence on unresolved scales demand the use of highly accurate numerical methods and massive computations [22-24]. In theory, we have to elaborate new approaches to study non-equilibrium, multi-scale, nonlinear and non-local dynamics, identify universal properties of asymptotic solutions, capture symmetries of RT flows, and account for their noisiness [8,15,25-33].

Despite these challenges, significant success has been recently achieved in the understanding of RTI and RT mixing [8,15-33]. It has been found that RT mixing with constant acceleration may exhibit order and has stronger correlation, weaker fluctuations and steeper spectra when compared to canonical turbulence [1-4,8,16-19,32]. Inspired by this success, we explore here an aspect of RT problem, which is critically important in realistic environments: the effect of variable acceleration on symmetries and invariants of RT mixing, and on its scaling, correlations and spectra [5,8,34-36]. We apply group theory and dimensional method [4,8,25,32,35-37] to keep the analysis from being too formal and too empirical, elaborate the concept of the modified rate of loss of specific momentum, discover the special class of self-similar solutions, and identify properties of RT mixing with variable acceleration that have not been discussed before. We find that in a broad range of the acceleration parameters dynamics of RT mixing can vary from super-ballistic to sub-diffusive and yet retains memory of deterministic conditions. Our results explain existing experiments, broaden horizons of studies of RT-relevant processes in nature and technology, and outline conditions at which canonical turbulence to be realized in RT mixing with variable acceleration. Supporting Information is provided for the corresponding sections [SI Appendix].

**Section - Theoretical approach**

*Governing equations.* RT dynamics is governed by the conservation of mass, momentum and energy [4,6-8]. In continuous approximation, for ideal fluids the governing equations are:

$$\partial\rho/\partial t + \partial\rho v_i/\partial x_i = 0,\ \partial\rho v_i/\partial t + \sum_{j=1}^{3}\partial\rho v_i v_j/\partial x_j + \partial P/\partial x_i = 0,\ \partial E/\partial t + \partial(E+P)v_i/\partial x_i = 0$$

where $(x_1, x_2, x_3, t) = (x, y, z, t)$ are the spatial coordinates and time, $(\rho, \mathbf{v}, P, E)$ are the fields of density, velocity, pressure and energy, $E = \rho(e + \mathbf{v}^2/2)$ and $e$ is specific internal energy. Gravity $\mathbf{g} = (0,0,-g), g = |\mathbf{g}|$ is directed from the heavy to the light fluid [4]. These nonlinear partial differential equations are augmented with the boundary value problem represented by a sub-set of nonlinear partial differential equations at a nonlinear freely evolving interface and by conditions at the outside boundaries of the domain [4,8]. They are further augmented with the initial value problem including initial perturbations of the flow fields at the interface and in the bulk. The boundary value problem is influenced by singularities and secondary instabilities; the initial value problem is ill-posed [6-8,25-35]. In spatially extended systems the flow can be periodic in the $(x,y)$ plane normal to the $z$ direction of gravity and is free of mass sources [8,25]. For non-ideal fluids, governing equations are further modified: with kinematic viscosity $\nu$, momentum equation is augmented with term $\left(-\rho\nu\,\partial^2 v_i/\partial x_j^2\right)$ [4].

Since the time of first hypotheses on existence of self-similar RT mixing and first endeavors to observe it in experiments and simulations, tremendous efforts have been undertaken to understand RT dynamics [8,15-30]. Significant success has been achieved by linear and weakly nonlinear theories in the quantifying the growth of RTI with constant and impulsive accelerations for various initial conditions; by interpolation models in the describing observational data of late time RT dynamics with nearly same sets of parameters; and by turbulence models in the estimating turbulence effect on RT mixing with the use of canonical scaling laws [6-8,15,25-34].

It is worth noting that while RT dynamics is a complex process with many coupled scales, it has some features of universality and order at early and at late times [6-8,15-25]. It can hence be studied from the first principles by applying group theory [8,15,25]. For linear and nonlinear RTI with constant and impulsive accelerations, group theory employs theory of discrete groups; its key results are symmetries of RT flows, multi-scale character of nonlinear dynamics (to which the coherent structure' wavelength and amplitude contribute), and tendency to keep isotropy in the plane normal to gravity [8,15,25,31]. For RT mixing with constant acceleration, group theory is implemented in the momentum model that has the same scaling symmetries as the governing equations; its key result - RT mixing may exhibit order - self-consistently explains a broad set of experiments [8,15-20,25,32,33].

While theory of RT dynamics illustrates what a real complexity is, in realistic environment the problem can be even more challenging, since RTI and RT mixing is often driven by variable acceleration [5,11-15]. Here we consider accelerations that are power-law functions of time. On the side of fundamentals, power laws can lead to special scaling properties of the dynamics [4,8,37]. On the side of applications they can be used to adjust the acceleration's time-dependence in observations [11-15]. Group theory analysis of the boundary value problem of linear and nonlinear RTI with variable acceleration is given in [36]. Here we focus on the analysis of invariants, scaling laws, correlations and fluctuations of RT mixing by implementing group theory in the momentum model and by further applying scaling transformations, Lie groups, and dimensional analysis.

*Momentum model*. The dynamics of a parcel of fluid undergoing RT mixing is governed by a balance per unit mass of the rate of momentum gain, $\mu$, and the rate of momentum loss, $\widetilde{\mu}$, as

$$\dot{h}=v,\ \dot{v}=\widetilde{\mu}-\mu$$

where $h$ is the length scale along the acceleration $\mathbf{g}$, $v$ is the corresponding velocity, $\widetilde{\mu}(\mu)$ is the magnitude of the rate of gain (loss) of specific momentum in the acceleration direction [1-4,8,11,15,32-36]. The rate of gain (loss) of specific momentum is $\widetilde{\mu} = \widetilde{\varepsilon}/v$ ($\mu = \varepsilon/v$), with $\widetilde{\varepsilon}(\varepsilon)$ being the rate of gain (loss) of specific energy. The rate of energy gain is $\widetilde{\varepsilon} = fgv$, $f = f(A)$, re-scaled $gf \to g$ hereafter; the rate of energy loss is $\varepsilon = Cv^3/L$, where $L$ is the length scale for energy dissipation and $C$ is drag. These lead to $\widetilde{\mu} = g$ and $\mu = Cv^2/L$.

The momentum model has the same symmetries and scaling transformations as the governing equations. Thus it captures the asymptotic behaviors, and invariant and scaling properties of the dynamics. For certain acceleration patterns, it can be solved by applying the Lie groups. We consider acceleration which is a power-law function of time, $g = Gt^a$, with exponent $a \in (-\infty, +\infty)$ and pre-factors $G > 0$ with dimensions $[G] = m/s^{(2+a)}$ and $[a] = 1$. We refer to the dynamics as to RT-type when it is driven by the acceleration, and as to RM-type when the acceleration effect is negligible asymptotically. We mark these types sub-scripts $RT, RM$ [6-10,34-36].

In RT flows, two macroscopic length scales contribute to the dynamics: the wavelength $\lambda$ in the normal plane and the amplitude $h$ in the direction of gravity $\mathbf{g}$, or the horizontal and the vertical scales [8,15,25,31-33]. The former is set by some deterministic conditions, e.g., by the wavelength of the initial perturbation or the mode of fastest growth [26,27]. The latter is regarded as an integral scale; it is believed to grow self-similarly in the mixing regime [8,15,32-34]. The scale for energy dissipation can be horizontal ($L \sim \lambda$, $\lambda \sim const$) or vertical ($L \sim h$), or be a combination of scales, $L \sim L(h, \lambda)$. Scale $L \sim \lambda$ corresponds to linear and nonlinear RTI (RMI) and scale $L \sim h$ - to RT (RM) mixing [8,15,32-36]. Time is $t > t_0 > 0$ with $t_0 > \{\tau_G, \tau_0\}$ and time-scales $\tau_G = (\lambda/G)^{1/(a+2)}$ and $\tau_0 = (\lambda/v_0)$ set by the acceleration $G$ and by initial growth-rate $v_0$. Drag $C$ is free parameter, $C \in (0, \infty)$. RT/RM linear, nonlinear and mixing dynamics may have their own values $C$ due to their distinct symmetries.

*Asymptotic dynamics.* In linear and nonlinear regimes with $L = \lambda$, for $a > -2$ the dynamics is RT-type, and the early-time and late-time asymptotic solutions are $v_{RT} = (\lambda/\tau_G)(t_0/\tau_G)^a((t-t_0)/\tau_G)$ with $|\dot{v}| \sim |\widetilde{\mu}|, |\mu/\widetilde{\mu}| \to 0$ and $v_{RT} = (\lambda/\tau_G)(t/\tau_G)^{a/2}C^{-1/2}$ with $|\mu| \sim |\widetilde{\mu}|, |\dot{v}/\mu| \to 0$. For $a < -2$ the dynamics is RM-type, and the early-time and late-time asymptotic solutions are $v_{RM} = v_0 - Cv_0(t-t_0)/\tau_0$ and $v_{RM} = (\lambda/C)/t$ with $|\dot{v}| \sim |\mu|, |\widetilde{\mu}/\mu| \to 0$. At $a = -2$ the solutions transit from RT- to RM-type with decrease of $(G/\lambda)$, in consistency with group theory analysis of the boundary value problem [36].

In mixing regime with $L = h$, for $a \geq a_{cr}$ the dynamics is RT-type and asymptotic solutions are $h_{RT} = B_{RT} t^{2+a}$, with algebraic imbalance of the rates of momentum $\tilde{\mu} \neq \mu, |\tilde{\mu}| \sim |\mu| \sim |\dot{v}| \sim t^a$. Their exponent is set by the acceleration's exponent, $(2+a)$, and pre-factor $B_{RT}$ - by the acceleration parameters and drag [34-36]. Solution $h_{RT}$ has the invariant $h/t^{(a+2)}$ due to the scaling symmetry of the governing equations $t \to e^\gamma t, h \to e^{\beta\gamma} h$ with transformation parameter $\gamma$ and constant $\beta = a + 2$. For $a \leq a_{cr}$, the dynamics is RM-type and asymptotic solutions are $h_{RM} = B_{RM} t^{2+a_{cr}}$, with asymptotic imbalance of the rate of momentum $|\tilde{\mu}/\mu| \to 0, |\mu| \sim |\dot{v}| \sim t^{a_{cr}}$. Their exponent is set by drag, $a_{cr} = -2 + (1+C)^{-1}$, and pre-factor $B_{RM}$ - by deterministic conditions. Solution $h_d$ has a point symmetry due to the arbitrary time of origin [34,52]. The mixing transits from RT to RM-type for $a \sim a_{cr}$, with critical exponent $a_{cr} \to -1$ for $C \to 0$ and $a_{cr} \to -2$ for $C \to \infty$ [33,51] [36]. While RT and RM type solutions should be coupled as particular and homogeneous solutions of the nonlinear equations, they are effectively de-coupled due to their distinct symmetries [8,15,32-36].

For $a > a_{cr}$ ($a < -2$) linear, nonlinear and mixing dynamics are RT (RM) type; for $-2 < a < a_{cr}$ linear and nonlinear dynamics is RT-type and mixing is RM-type [36]. RT mixing may develop for $a > a_{cr}$ due to algebraic imbalance of the rates of momentum $\tilde{\mu} \neq \mu$, when the scale for energy dissipation is the vertical scale $L \sim h$, and/or when the scale for energy dissipation is the horizontal scale $L \sim \lambda$ increasing as $\lambda \sim G t^{a+2}$. RM mixing may develop for $a < a_{cr}$ due to asymptotic imbalance of the rates of momentum $|\tilde{\mu}/\mu| \to 0$, when the scale for energy dissipation is the vertical scale $L \sim h$, and/or when $L \sim \lambda$ increasing as $\lambda \sim t^{(a_{cr}+2)+(a_{cr}-a)}$ ($\lambda \sim t^{(a_{cr}+2)}$) for $-2 < a < a_{cr}$ ($a < -2$). The growth of horizontal scale is possible (and not necessary) condition for self-similar mixing to occur, in agreement with group theory analysis of RT (RM) structures and transitions [8,15,25,36,38].

**Section - RT mixing with variable acceleration**

To study scaling laws, correlation and spectra of self-similar RT (RM) mixing with variable acceleration, we employ the scaling transformations and dimensional analysis as in the classical works [4,37].

***Dynamic properties.*** In RT-type mixing with $a > a_{cr}$, the length scales with time as $L \sim t^{a+2}$, and the velocity - as $v \sim t^{a+1}$. Length scale increases with time for any $a > a_{cr}$, whereas velocity scale increases with time for $a > -1$, is constant at the steady flex point $a = -1$, and decreases with time for $a < -1$. By comparing with scaling laws for diffusion $L \sim t^{1/2}$ and ballistics $L \sim t^2$, and by marking similar/quicker/ slower processes with prefixes quasi/super/sub, we find that for RT-type mixing with $a > a_{cr}$ the dynamics $L \sim t^{a+2}$ is super-ballistic for $a > 0$, ballistic at $a = 0$, super-diffusive for $0 > a > -3/2$, quasi-diffusive at $a = -3/2$, and sub-diffusive for $-3/2 > a > a_{cr}$ [2,4,36,37,39]. At the steady flex point $a = -1$, a transition occurs from the dynamics with larger velocities at larger scales to that with larger velocities at smaller scales. In RM-type mixing $a < a_{cr}$, length scale increases with time $L \sim t^{(a_{cr}+2)}$, velocity scale decreases with time $v \sim t^{(a_{cr}+1)}$, larger velocities correspond to smaller scales, and the dynamics is sub-diffusive for $C > 1$, Figure 1 [36]. Canonical turbulence is super-diffusive, $L \sim t^{3/2}$, with larger velocities at larger scales [1-4,37,39-41].

***Symmetries, scaling transformations and invariants.*** RT mixing and canonical turbulence have symmetries in statistical sense. For isotropic and homogeneous turbulence these are Galilean transformations, translations in time, as well as translations, rotations and reflections in space [2,4]. In the governing equations, incompressible turbulent dynamics is invariant with respect to scaling transformation, $L \to LK$, $v \to vK^n$, $t \to tK^{1-n}$. In the limit of vanishing viscosity, $\nu/vL \to 0$, with $\nu \to \nu K^{1+n}$, the exponent is $n = 1/3$ and the invariant is the rate of energy dissipation, $\varepsilon = \sum_{i,j} \nu (\partial v_i / \partial x_j)^2$, with $\varepsilon \sim v^3/L$ and $\varepsilon \to \varepsilon K^{3n-1}$ [1-4].

RT mixing is non-inertial, and is invariant with respect to translations, rotations and reflections in the plane normal to the acceleration **g**. It is invariant with respect to scaling transformation, $L \to LK$, $t \to tK^{1-n}$, $v \to vK^n$. For constant acceleration, $a = 0$ and $g = Gt^0 = g_0$, with $g_0 \to g_0 K^{2n-1}$, in the limit of vanishing viscosity, $\nu/vL \to 0$, with $\nu \to \nu K^{1+n}$, the transformation exponent is $n = 1/2$ and the invariant is the rate of loss of specific momentum in the direction of gravity, $\mu$, or $\mu \sim v^2/L$, with $\mu \to \mu K^{2n-1}$, where the components are $\mu_i = \sum_j \nu (\partial^2 v_i / \partial x_j^2)$ of the vector of the rate of momentum loss. The exponent $n = 1/2$ leads to invariance of $\tilde{\mu}, \mu$, whereas the rates of energy gain and energy dissipation are time-dependent, $\tilde{\varepsilon}, \varepsilon \sim g_0^3 t$ [8,15,32].

For variable acceleration, $g = Gt^a$, in RT-type mixing with $a > a_{cr}$, the dynamics is invariant with respect to scaling transformation $L \to LK$, $v \to vK^n$, $t \to tK^{1-n}$, conditional that $\nu \to \nu K^{1+n}$ and $G \to GK^{n(a+2)-(a+1)}$. In this case in the limit of vanishing viscosity, $\nu/vL \to 0$, the rates of loss and gain of energy $\varepsilon(\tilde{\varepsilon})$ and momentum $\mu(\tilde{\mu})$ are scale-dependent. The exponent $n = (a+1)/(a+2)$ leads to invariance of $M = \mu/t^a$, which we refer as to modified rate of momentum loss, $M \sim v^{(a+2)}/L^{(a+1)}$, and the invariance of modified rate of momentum gain $\tilde{M} = \tilde{\mu}/t^a$, $\tilde{M} = G$. The value $M(\tilde{M})$ can be viewed as a fractional $a$ th order time-derivative $\partial^a/\partial t^a$ of the rate of loss (gain) of specific momentum in the direction of gravity [8,15,32,39]. In RM-type mixing with $a < a_{cr}$, the invariant is $M_{cr} = \mu/t^{a_{cr}}$, $M_{cr} \sim v^{(a_{cr}+2)}/L^{(a_{cr}+1)}$ with $a_{cr}$ set by drag; $\tilde{M}_{cr} = \tilde{\mu}/t^{a_{cr}}$ is zero asymptotically $\tilde{M}_{cr} \to 0$ [36].

In RT (RM) type mixing with $a > a_{cr}$ ($a < a_{cr}$) the invariance of $M$ ($M_{cr}$) implies that the modified rate of momentum is constant, and there is the momentum transport between the scales in real space [8,15]. In canonical turbulence, the invariance of energy dissipation rate $\varepsilon \sim v^3/L$ is compatible with existence of inertial interval and non-dissipative energy transfer between the scales in wave-vector space [1-4]. Enstrophy and helicity are other invariants of canonical turbulence [1-4]. In RT mixing with variable acceleration these values are scale-dependent. In RT-type mixing with $a > a_{cr}$, the invariant $M \sim v^{(a+2)}/L^{(a+1)}$ leads to $\varepsilon \sim L^{(2a+1)/(a+2)}$ and $\varepsilon \sim t^{2a+1}$. The rate of energy dissipation increases (decreases) with time for $a > -1/2$ ($a < -1/2$) and is constant at $a = -1/2$. The rate of momentum loss, $\mu \sim L^{a/(a+2)}$ and $\mu \sim t^a$, increases (decreases) with time for $a > 0$ ($a < 0$) and is constant at $a = 0$. In RM-type mixing with $a < a_{cr}$, rates of energy dissipation and momentum loss decay with time for any $C$, $\varepsilon \sim t^{2a_{cr}+1}$, $\mu \sim t^{a_{cr}}$.

***Scaling laws and correlations.*** In RT-type mixing $a > a_{cr}$, invariant $M \sim v^{(a+2)}/L^{(a+1)}$ leads to $v^{(a+2)}/L^{(a+1)} \sim v_l^{(a+2)}/l^{(a+1)}$, where $v_l$ is a velocity at a length scale $l$. This results in the scaling for velocity $v_l/v \sim (l/L)^{(a+1)/(a+2)}$ and the $N$ th order structure function $\sim (lM)^{N(a+1)/(a+2)}$. In canonical turbulence the scaling for velocity is $v_l/v \sim (l/L)^{1/3}$ and for the $N$ th order structure function is $\sim (l\varepsilon)^{N/3}$ due to invariance of $\varepsilon \sim v^3/L$. By comparing these scaling exponents we find that in RT-type mixing the velocity correlations are stronger (weaker) for $a > -1/2$ ($a < -1/2$) when compared to turbulence, and have turbulent scaling laws at $a = -1/2$. In RM-type mixing $a < a_{cr}$, invariant $M_{cr} \sim v^{(a_{cr}+2)}/L^{(a_{cr}+1)}$ leads to scaling for velocity $v_l/v \sim (l/L)^{(a_{cr}+1)/(a_{cr}+2)}$ and the $N$ th order structure function $\sim (lM_{cr})^{N(a_{cr}+1)/(a_{cr}+2)}$, with larger velocities at smaller scales, Figure 1.

In RT-type mixing $a > a_{cr}$, the Reynolds number $Re = vL/\nu$ is scale-dependent, $Re \sim G^2 t^{(2a+3)}/\nu$. It increases (decreases) with time for $a > -3/2$ ($a < -3/2$). Defining local Reynolds number as $Re_l = v_l l/\nu$, we find the Reynolds number scaling $Re_l/Re \sim (l/L)^{(2a+3)/(a+2)}$, the viscous scale $l_\nu \sim (\nu^{(a+2)}/M)^{1/(2a+3)}$, and the span of scales, $L/l_\nu \sim t^{a+2}(M^2/\nu)^{(a+2)/(2a+3)}$. In RT-type mixing the span of scales increases with time, and the viscous scale is finite. It is set by viscosity and acceleration and is comparable with the fastest growing wavelength $\sim (\nu^{(a+2)}/G)^{1/(2a+3)}$. In RM-type mixing with $a < a_{cr}$ the scaling is $Re_l/Re = (l/L)^{(2a_{cr}+3)/(a_{cr}+2)}$ for the Reynolds number, $l_\nu \sim (\nu^{(a_{cr}+2)}/M_{cr})^{1/(2a_{cr}+3)}$ for the viscous scale is, and $L/l_\nu \sim t^{a_{cr}+2}/l_\nu$ for the span of scales. While in late-time RM-type mixing the viscous scale is finite, it is distinct from that at early times $\sim (\nu/v_0)$. In turbulence, $Re_l/Re \sim (l/L)^{4/3}$, $l_\nu \sim (\nu^3/\varepsilon)^{1/4}$, and $L/l_\nu \sim L(\varepsilon/\nu^3)^{1/4}$, Figure 1 [1-4,8,15,32].

***Properties of fluctuations.*** Fluctuations are essential for turbulent systems; their strength is expected to exceed the noise from deterministic conditions. In canonical turbulence the invariance of energy dissipation rate $\varepsilon \sim v^3/L$ leads to diffusion scaling law for velocity fluctuations $v \sim t^{1/2}$ [4,39]. For RT-type mixing with $a > a_{cr}$ the invariance of modified rate of momentum $M$ with $[M] = m/s^{(a+2)}$ leads to scaling for velocity fluctuations $v \sim t^{(a+1)}$. These fluctuations are super-diffusive for $a > -1/2$, quasi-diffusive at $a = -1/2$, and sub-diffusive for $-1/2 > a > a_{cr}$. For RM-type mixing with $a < a_{cr}$ the invariant is $M_{cr}$, and velocity fluctuations scaling is $v \sim t^{(a_{cr}+1)}$.

To compare the strengths of velocity fluctuations due to turbulence and due to deterministic conditions, consider two parcels of fluids entrained in the flow with a time-delay $\tilde{\tau}$. Canonical turbulence is a stochastic process with self-generated fluctuations, and fluctuations due to turbulence $\sim (\varepsilon v \tilde{\tau})^{1/3}$ are substantially stronger than the parcels' relative velocity $\sim (\varepsilon \tilde{\tau})^{1/2}$ [4]. In RT (RM) type mixing with $a > a_{cr}$ ($a < a_{cr}$), the 'turbulent' fluctuations $\sim M\tilde{\tau}^{(a+1)}$ ($M_{cr}\tilde{\tau}^{(a_{cr}+1)}$) are comparable to the parcels' relative velocity $\sim M\tilde{\tau}^{(a+1)}$ ($M_{cr}\tilde{\tau}^{(a_{cr}+1)}$). Besides, the ratio of fluctuating and mean velocities is $\sim (\tilde{\tau}/t)^{(a+1)}$ ($(\tilde{\tau}/t)^{(a_{cr}+1)}$) [8,15,32]. Hence, in RT (RM) type mixing, fluctuations are sensitive to deterministic conditions. In RT-type mixing their strength decays with time for $a > -1$, is time-independent at $a = -1$, and increases with time for $-1 > a > a_{cr}$; in RM-type mixing it increases with time.

In statistically steady canonical turbulence, the invariance of energy dissipation rate leads to kinetic energy spectrum $E(k) \sim \varepsilon^{2/3} k^{-5/3}$ $(E(\omega) \sim \varepsilon \omega^{-2})$, where $E(k)(E(\omega))$ is the spectral density and $k(\omega)$ is the wavevector (frequency) [2,4]. In RT-type mixing with $a > a_{cr}$ an accurate determination of spectra is a challenge since the dynamics is statistically unsteady [8,15,32]. Dimensional analysis suggests that due to the invariance of the modified rate of momentum $M$, the spectral density of the specific kinetic energy may have the form $E(k) \sim M^{2/(a+2)} k^{-(3a+4)/(a+2)}$ $(E(\omega) \sim M^2 \omega^{-(2a+3)})$. The spectrum is steeper (more gradual) than Kolmogorov's spectrum for $a > -1/2$ $(a_{cr} < a < -1/2)$ and has exponent $-5/3$ $(-2)$ at $a = -1/2$, Figure 1. For $a_{cr} < a < -1$, scaling exponent in $E(k)(E(\omega))$ may change sign, since, for $a < -1$ larger velocities correspond to smaller scales, in contrast to super-diffusive turbulence, Figure 1. In RM-type mixing with $a < a_{cr}$, $E(k) \sim M_{cr}^{2/(a_{cr}+2)} k^{-(3a_{cr}+4)/(a_{cr}+2)}$ $(E(\omega) \sim M_{cr}^2 \omega^{-(2a_{cr}+3)})$. Signs of scaling exponents are consistent with sub-diffusive character of RM-type mixing, Figure 1.

**Section – Special class of self-similar dynamics**

Consider properties of RT mixing in a broad range of the acceleration exponent, $a \in (-\infty, \infty)$, Figure 1.

***Super-ballistics.*** For $a > 0$ RT-type mixing is super-ballistic and correlations are strong. For $a \to \infty$, the invariant is $M \sim v/L$, the velocity and Reynolds number scale as $v_l/v \sim (l/L)$ and $Re_l/Re \sim (l/L)^2$, and the spectral density is $E(k) \sim k^{-3}$. At $a = 0$, the invariant is $M = \mu \sim v^2/L$, the velocity and Reynolds number scale as $v_l/v \sim (l/L)^{1/2}$ and $Re_l/Re \sim (l/L)^{3/2}$, and the spectral density is $E(k) \sim k^{-2}$. This dynamics is ballistic, with strong correlations and weak fluctuations whose strength is set by deterministic conditions and decays with time $\sim t^{-1}$. For $a < 0$, RT-type mixing is sub-ballistics; for $-1 < a < 0$, it has larger velocities at larger scales. For $a > -1/2$, RT-type mixing is super-Kolmogorov. Its correlations are stronger and spectra are steeper than those in turbulence, suggesting that strongly accelerated RT mixing may laminarize, Figure 1 [8,15, 42,43].

***Quasi-Kolmogorov.*** At $a = -1/2$, RT-type mixing is quasi-Kolmogorov since it has the same scaling properties as canonical turbulence, including the invariance of energy dissipation rate $\varepsilon \sim v^3/L$, the velocity and Reynolds number scale as $v_l/v \sim (l/L)^{1/3}$ and $Re_l/Re \sim (l/L)^{4/3}$, and spectral density is $E(k) \sim k^{-5/3}$ [4]. Yet, in RT-type mixing fluctuations are sensitive to deterministic conditions, and their strength decays with time $\sim t^{-1/2}$. For $-1 < a < -1/2$ in RT-type mixing correlations are weaker and spectra are more gradual than in canonical turbulence, Figure 1.

*Steady flex.* At $a = -1$ RT-type mixing is steady, $L \sim t$, its velocities are scale-independent $v_l \sim v$, and the Reynolds number scales as $Re_l/Re = (l/L)$. At this steady flex point, velocity fluctuations are set by deterministic conditions, their strength is constant, and spectral density is $E(k) \sim k^{-1}$. For $a < -1$ RT-type mixing has larger velocities at smaller scales. Yet for $-3/2 < a < -1$ the dynamics is super-diffusive, Figure 1.

*Quasi-diffusion.* At $a = -3/2$ RT-type mixing is quasi-diffusive, with diffusion scaling law for length $L \sim t^{1/2}$ and velocity $v_l/v \sim (l/L)^{-1}$, and scale-invariant Reynolds number $Re_l \sim Re$. At $a = -3/2$ fluctuations are set by deterministic conditions, and their strength increases with time $\sim t^{1/2}$ leading to spectral density $E(k) \sim k$. The sign of spectral exponent is consistent with quasi-diffusive character of the dynamics having larger velocities at smaller scales. For $a_{cr} < a < -3/2$, RT-type mixing is sub-diffusive. The smaller the acceleration exponent the stronger the effect of deterministic conditions is, Figure 1.

*Sub-diffusion.* At $a \sim a_{cr}$, $a_{cr} = -2 + (1+C)^{-1}$, the mixing becomes RM-type. For $-\infty < a < a_{cr}$ RM-type mixing is set by drag and deterministic conditions. As a sub-diffusive process, RM-type mixing has intense motions at small scales with larger velocities at smaller scales. The velocity and Reynolds number scale as $v_l/v \sim (l/L)^{-C}$ and $Re_l/Re \sim (l/L)^{1-C}$, the strength of deterministic fluctuations increases with time $\sim t^{(C+1)/C}$, and the spectral density is $E(k) \sim k^{2C-1}$, Figure 1.

**Section – RT mixing in nature, technology, laboratory**

Our results explain and agree with existing observations and serves to better understand RT-relevant phenomena in nature and technology.

*Supernovae.* Supernovae are a central problem in astrophysics due to their role in stellar evolution and nuclear synthesis [12]. Supernova's explosion is driven by a blast-wave causing intensive RT/RM mixing of materials of a progenitor star and enabling conditions for synthesis of heavy mass elements [12,36]. Blast-wave induced acceleration is a power-law function of time, since blast waves are described by self-similar solutions, whose invariant $F, [F] = m/s^\theta$, $0 < \theta < 1$, is a function on energy release, density and pressure [36]. For the power-law solution, the length scale is $L \sim t^\theta$ and acceleration is $\sim t^{\theta-2}$ [36,37]. By comparing exponents $(\theta - 2)$ and $a_{cr}$, we find that blast-wave-driven mixing can be RT-type with $a \sim a_{cr}$ for $C < (\theta^{-1} - 1)$ and RM-type with $a < a_{cr}$ for $C > (\theta^{-1} - 1)$. The mixing is sensitive to deterministic conditions and has larger velocities at smaller scales (thus explaining anisotropy and richness of structures in supernovae remnants). Our theory further suggests that in such dynamics a super-diffusive turbulence may be a challenge to occur. Mechanisms other than turbulence may enable transport and accumulation of energy at small scales needed for nuclear synthesis. These may include localizations and trapping (typical for sub-diffusive processes), in excellent agreement with observations of small-scale non-uniform structures in RM flows [36,38].

*High energy density laboratory plasmas.* The conditions of high energy density in supernovae can be achieved at high power laser facilities [18,19,35,44]. Our analysis is in excellent agreement with experiments designed to mimic blast-wave-driven RT mixing in laboratory plasmas [18,35,44]. These experiments find that RT mixing has high degree of coherence. Its late-time dynamics is sensitive to initial (deterministic) conditions, is dominated by the vertical scale, and also exhibits some small-scale structures. In the experiments the acceleration exponent and the Reynolds number are $a \sim -1, Re \sim 10^6$ [18,35]. Our analysis achieves excellent agreement with the observed properties of the dynamics, including tendencies of RT mixing to keep order, retain memory of deterministic conditions, be dominated by the vertical scale, and exhibit intense small-scale structures due to scale-invariance of velocity fluctuations at $a = -1$, Figure 1 [18,35]. Our results can be further applied for better control fluid instabilities in high energy density plasmas, including inertial confinement fusion [13,44].

*Experiments in fluids.* Our theory finds that RT mixing dynamics varies from super-ballistic to sub-diffusive depending on the acceleration and is sensitive to deterministic conditions for any acceleration. This result is in excellent agreement with experiments [16,17]. The experiments are conducted in gas-gas and gas-liquid system, and apply strong and weak shocks, compressed air as well as gas mixture detonation to strongly accelerate the fluids and achieve self-similar regimes with high Reynolds numbers $Re \sim 3.2 \times 10^6$ [15-17]. To control initial perturbations in gas-liquid system, experiments apply the jelly technique, by using jellies of small concentration of aqueous gelatin solutions with zero plasticity and weak strength behaving as incompressible fluids under strong pressure [16,17]. The experiments unambiguously find that in a broad parameter regime (including periodic and localized perturbations of the interface and the flow fields, convergent and spatially extended geometries, steady and variable accelerations) RT mixing is interfacial, heterogeneous and anisotropic; it retains memory of deterministic conditions and may keep order [15-17]. Our analysis achieves excellent agreement with these experiments.

*Laminarization of accelerated flows.* Our theory finds that in RT-type mixing with acceleration exponents $a > -1/2$ the correlations are stronger and spectra are steeper than in canonical turbulence. This suggests that strongly accelerated RT mixing may laminarize. Our results are consistent with classical experiments in accelerated flows, including flows in curved pipes and accelerated boundary layers [42,43]. The experiments find that turbulent flows may laminarize when they are sharply accelerated. In our analysis sharp accelerations correspond to large acceleration exponents, thus agreeing with the observations [42,43].

*Interplay of acceleration and turbulence.* Our theory self-consistently explains experiments and simulations of RT mixing with constant acceleration which found spectra steeper than Kolmogorov [20]. It also explains the flattening of initially steep spectra in simulations of RM mixing [24]. Our results can be further applied to study the interplay of turbulence and acceleration. According to our theory RT mixing is quasi-Kolmogorov at the acceleration exponent $a = -1/2$ since it has the same scaling laws, correlations and spectra as canonical turbulence [4]. May this imply that variable acceleration $g \sim t^{-1/2}$ may produce and maintain turbulence? Our theory suggests: Yes, it may, conditional there is also a source supplying turbulent energy to the system at a constant rate [1-5], and the flow is being homogenized [45,46]. While the acceleration may quickly set proper scaling laws, high Reynolds numbers and large spans of scales, the source of turbulent energy is needed to keep fluctuations strong and stochastic [1-4].

*Anomalous scaling.* Our analysis finds significant departures of invariant, scaling, spectral and correlation properties of self-similar RT mixing from those of canonical turbulence. In realistic turbulent processes, departures from Kolmogorov's scaling laws are found in isotropic homogeneous turbulence, turbulent boundary layers, passive scalar mixing, buoyancy-driven turbulence, turbulent convection, compressible turbulence [3,40,47-50]. These departures are referred to as to 'anomalous scaling'. Our analysis identifies the special class of self-similar solutions whose dynamics can vary from super-ballistics to sub-diffusion depending on a single dimensionless parameter. It suggests that the 'anomalous scaling' can serve to systematically incorporate conditions of realistic environments and to better understand complex and turbulent processes in nature and technology [3,5,8,40,47-50].

**Section - Discussion**

Rayleigh-Taylor mixing is the generic problem we inevitably encounter when expanding our knowledge of turbulence and when studying a broad range of processes in nature and technology, from supernovae and fusion to fossil fuel recovery and overturned cup of water [1-45]. In this work we have focused on the effect of variable acceleration on self-similar RT mixing, and on its scaling, correlations and spectra.

In full consistency with classic approaches [4,37] we have applied group theory and dimensional analysis to study properties of RT dynamics with power-law time-dependent acceleration. We have elaborated the concept of invariance of modified rate of loss of momentum, discovered the special class of self-similar solutions, and identified properties of RT mixing that have not been discussed before. We have found that dynamics of self-similar RT mixing can vary from super-ballistics to sub-diffusion depending on the acceleration's exponent and retains memory of deterministic conditions for any exponent. For super-ballistic, ballistic and super-Kolmogorov dynamics, RT mixing has strong correlations and weak fluctuations which may lead to order laminarization. For quasi-Kolmogorov dynamics, in addition to proper scaling laws set by the acceleration, a source of turbulent energy is needed for keeping fluctuations strong and stochastic. For quasi- and sub-diffusive dynamics in RT/RM – type mixing larger velocities correspond to smaller scales, in consistency with classical results [4,37]. The smaller acceleration exponents the more sensitive the fluctuations are to deterministic conditions.

Our results achieve excellent agreement with a broad range of experiments and simulations in fluids and plasmas [15-20,42-44]. These include: experiments in fluids detected laminarization of strongly accelerated turbulent flows; experiments in plasmas observed high degree of coherence and dominance of the vertical scale in RT mixing; experiments in fluids unambiguously found strong sensitivity of RT mixing to deterministic conditions; experiments and simulations observed anomalous spectral properties of RT/RM. New experiments can be designed to quantify the interplay of acceleration, turbulence and deterministic conditions in RT mixing. For instance, one may employ narrow band initial conditions with various symmetries, and study the effect of interference of initial perturbation waves on order and disorder in RT mixing [25,38]. One can also employ a broad band incoherent perturbation with initially steep (gradual) spectra, and observe the evolution of spectral properties of RT mixing [20,24]. One can further vary acceleration parameters, and observe, diagnose and quantify RT dynamics in ballistic, quasi-Kolmogorov, steady flex, super-diffusive, and sub-diffusion regimes [8,15,36,39].

Our results serve for better understanding and control of RT-relevant phenomena in nature and technology. For instance, for supernovae, our analysis explains the richness of structures observed in supernovae remnants and reveals the mechanism of energy localization at small scales for nuclear synthesis [12,36]. For fusion, strong sensitivity of RT mixing to deterministic conditions suggests the opportunities for flow control of plasma flows by means of variable acceleration and deterministic conditions not explored before [13,44]. For turbulent flows, our analysis is fully consistent with experiments, simulations and theory of turbulent flows [3,40,47-50], and indicates that it is a highly formidable task to create canonical turbulence in a laboratory [1-5]. For fundamentals, the existence of special class of self-similar dynamics is directly linked to fractional analysis, dynamical systems, and boundary value problems [4,8,15,25,37,39,45,46], to be studied in the future. To conclude, the problem of water flowing from an overturned cup remains a source of inspiration for researchers in science, mathematics and engineering, and is well open for a curious mind.

**Acknowledgements**

The author thanks the University of Western Australia (AUS), the National Science Foundation (USA).

**Figure captions**

Figure 1: Scaling exponents in RT mixing with variable acceleration for velocity, Reynolds number and spectral density at some drag value.

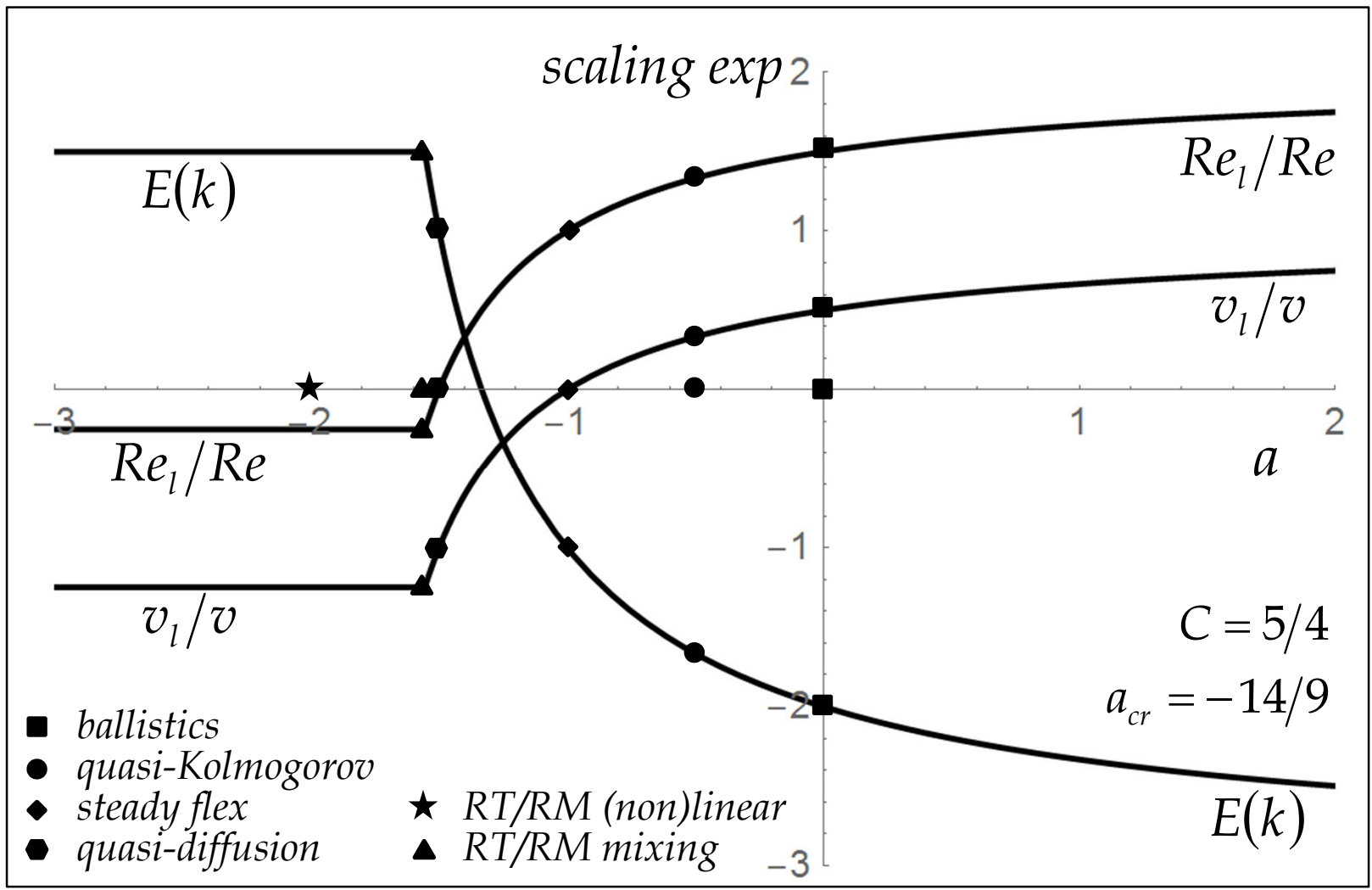